\documentclass[12pt]{article}
\usepackage{mathrsfs}
\usepackage{authblk}
\usepackage{tipa}
\usepackage[capposition=below]{floatrow}
\usepackage{graphics}
\usepackage[svgnames]{xcolor}
\usepackage{amsmath}
\usepackage{amssymb}
\usepackage{bm}
\usepackage{amsfonts}
\usepackage{ntheorem}
\usepackage[colorlinks=true, linkcolor=blue!60!black, citecolor=green!50!black, linktoc=all]{hyperref}
\definecolor{darkgreen}{rgb}{0,0.5,0}
\definecolor{darkblue}{rgb}{0,0,0.6}
\definecolor{purple}{rgb}{0.4,0.15,0.21}
\numberwithin{equation}{section}
\usepackage{feynmf}
\usepackage{extarrows}
\usepackage{slashed}
\usepackage{graphicx}
\usepackage[page,title,titletoc,header]{appendix}
\usepackage{indentfirst}
\usepackage{color}
\usepackage{cancel}
\author[]{\normalsize{\textbf{Bart Horn}}\thanks{bh2478@columbia.edu}}
\author[]{\normalsize{\textbf{Lam Hui}}\thanks{lh399@columbia.edu}}
\author[]{\normalsize{\textbf{Xiao Xiao}}\thanks{xx2146@columbia.edu}}
\affil[]{\normalsize{\emph{Physics Department and Institute for Strings, Cosmology and Astroparticle Physics,}}\\
\normalsize{\emph{Columbia University 
, New York, NY, 10027, USA}}}
\begin{document}
\title{\LARGE{Lagrangian space consistency relation for large scale structure}}
\date{}
\maketitle
\abstract{Consistency relations, which relate the squeezed limit of an (N+1)-point correlation function to an N-point function, 
are non-perturbative symmetry statements that hold even if the
associated high momentum modes are 
deep in the nonlinear regime
and astrophysically complex. 
Recently, Kehagias \& Riotto and Peloso \& Pietroni discovered a consistency relation
applicable to large scale structure. We show that this can be recast
into a simple physical statement 
in Lagrangian space:\ that the squeezed correlation function (suitably normalized) vanishes.
This holds regardless of whether the correlation observables are at
the same time or not, and regardless of whether multiple-streaming is present.
The simplicity of this statement suggests that an analytic understanding of 
large scale structure in the nonlinear regime
may be particularly promising in Lagrangian space.}
\clearpage

\tableofcontents

\numberwithin{equation}{section}
\section{Introduction}%
\label{intro}

Consistency relations are statements which relate the squeezed limit of an (N+1)-point correlation function to an N-point function of cosmological perturbations;\ i.e.,\ they
take the following schematic form in momentum space:
\begin{eqnarray}
\label{schematic}
\lim_{{\bf k} \rightarrow 0} {\langle \pi_{\bf k} {\cal O}_{\bf k_1} {\cal O}_{\bf k_2} ... {\cal O}_{\bf k_N} \rangle^{c'}
\over P_\pi ({\rm k})} \sim \langle {\cal O}_{\bf k_1} {\cal O}_{\bf k_2} ... {\cal O}_{\bf k_N} \rangle^{c'}\, ,
\end{eqnarray}
where $\pi_{\bf k}$ represents a squeezed wavemode (long wavelength) of what turns out
to be a Goldstone boson or pion, $P_\pi ({\rm k})$ is the power spectrum of the pion 
(${\rm k}$ represents the magnitude of the vector ${\bf k}$), 
and ${\cal O}$ represents observables at high momenta
${\bf k_1}, ..., {\bf k_N}$. The symbol $\langle ... \rangle^{c'}$ denotes the connected
correlation function with the overall delta function removed.
Consistency relations can be understood as analogues of `soft-pion' theorems in particle physics, which arise generally when a symmetry is spontaneously broken/nonlinearly realized.  In the case of cosmology, the symmetries in question are diffeomorphisms (i.e.\ coordinate transformations), and consistency relations arise from a particular set of residual symmetries of a given gauge where the transformation does not fall off at infinity.  The first example of a consistency relation was pointed out
by Maldacena \cite{Maldacena:2002vr} in the context of a computation of
the three-point correlation function from inflation. The utility of this as a test
of single field/clock inflation was emphasized by Creminelli \& Zaldarriaga \cite{Creminelli:2004yq}. Recent work pointed out new symmetries and
therefore further consistency relations \cite{Creminelli:2012ed,Hinterbichler:2012nm},
indeed an infinite tower of them \cite{Hinterbichler:2013dpa},
and explicated their non-perturbative nature \cite{Assassi:2012zq, Assassi:2012et, Kehagias:2012pd, Goldberger:2013rsa, Pimentel:2013gza, Berezhiani:2013ewa}. 

These consistency relations are extremely robust:\ they remain valid
when the high momentum modes (${\cal O}$ in Eq.\ \ref{schematic}) are deep
in the nonlinear regime, and even when the observables are astrophysically
complex (such as galaxy density). This point might appear academic when
applied to (small) perturbations in the early universe, such as are revealed in the
cosmic microwave background. When applied to large scale structure (LSS)
in the late universe, however, the robustness of the consistency relations
becomes very interesting. 
It thus came as welcome news when Kehagias \& Riotto \cite{KR} and Peloso \& Pietroni \cite{PP} 
(KRPP) pointed out that non-trivial consistency relations exist even 
if all wavemodes (including the squeezed one) 
are within the Hubble radius, within the Newtonian regime
which is the realm of LSS
(see also \cite{Creminelli:2013mca,Peloso:2013spa,Kehagias:2013rpa,Berger:2014wta,Creminelli:2013poa,Valageas:2013cma,Creminelli:2013nua,us}).

The KRPP consistency relation can be stated in the following form:
\begin{eqnarray}
\label{KRPPEuler}
\lim_{{\bf k} \rightarrow 0}
{\langle v^j_{\bf k} (\eta) \, {\cal O}_{{\bf k}_1} (\eta_1)  \,
  ... \, {\cal O}_{{\bf k}_N} (\eta_N) \, \rangle^{c'} \over
P_v ({\rm k}, \eta)} = {i {\rm k}^j} \sum_{a=1}^N {D(\eta_a) \over D'(\eta)} 
{{\bf k} \cdot {{\bf k}_a} \over {\rm k}^2} \langle {\cal O}_{{\bf
    k}_1}  (\eta_1) \, .... \, {\cal O}_{{\bf k}_N} (\eta_N) \,
\rangle^{c'} \, ,
\end{eqnarray}
where $v^j_{\bf k}$ is the $j$-th component of the peculiar velocity
in momentum space, 
$P_v$ is the velocity power spectrum defined by
$\langle v^i_{\bf k}  (\eta) \, v^j_{{\bf k}'} {}^{*} (\eta)\rangle
= (2\pi)^3 \delta_D ({\bf k} - {\bf k}') ({\rm k}^i {\rm k}^j/{\rm
  k}^2)  P_v({\rm k}, \eta)$,\footnote{This form of the velocity power spectrum assumes no
vorticity. This is acceptable since $P_v ({\rm k})$ is used only for
small ${\rm k}$, or large scales, where the growing mode initial condition ensures gradient flow.}
the observables can be thought of as mass or galaxy overdensity at
different momenta and times, and $D$ and $D'$ represent
the linear growth factor and its conformal time derivative.
The fluctuation variables will in general depend on time, although we will often suppress the time dependence to simplify the notation: $v^j_{\bf k}$
(and its power spectrum) is at conformal time $\eta$,
${\cal O}_{{\bf k}_1}$ is at time $\eta_1$, and so on. The times need
not be equal.  The symbol ${\rm k}^2$ denotes ${\bf k} \cdot {\bf k}$.

We wish to show that the KRPP consistency relation takes a
particularly
simple form in Lagrangian space:
\begin{eqnarray}
\label{KRPPLag00}
\boxed{
\lim_{{\bf p} \rightarrow 0}
{\langle \pmb{v}_{\bf p} (\eta) \,  {\cal O}_{{\bf p}_1} (\eta_1) \,
  ... \, {\cal O}_{{\bf
      p}_N} (\eta_N) \, \rangle^{c'} \over
P_v ({\rm p}, \eta)} = 0 \, .}
\end{eqnarray}
Unless otherwise stated, we use ${\bf p}$ to denote momentum
in Lagrangian space and ${\bf k}$ to denote momentum in Eulerian space.
In other words:
\begin{eqnarray}
\label{Fourier}
{\cal O}_{\bf k} = \int d^3 {\bf x} \, {\cal O} ({\bf x}) e^{i {\bf k} \cdot
  {\bf x}} \quad , \quad
{\cal O}_{\bf p} = \int d^3 {\bf q} \, {\cal O} ({\bf x} ({\bf q})) e^{i {\bf p} \cdot
  {\bf q}}  \, ,
\end{eqnarray}
where ${\bf x}$ and ${\bf q}$ are the Eulerian space 
and Lagrangian space coordinates respectively.\footnote{The definitions given apply even in the presence of
multiple streaming. See discussion in \S \ref{notation}.}
In both cases, we rely on context to distinguish
between ${\cal O}$ in Fourier space and ${\cal O}$ in configuration space.

Since the 
velocity ${\pmb v}$ (whose $j$-th component is $v^j$) 
is nothing other than the time derivative of the displacement 
$\pmb{\Delta}$ in Lagrangian space, we can also rewrite the
Lagrangian space consistency relation as:
\begin{eqnarray}
\label{KRPPLag}
\boxed{
\lim_{{\bf p} \rightarrow 0}
{\langle \pmb{\Delta}_{\bf p} (\eta) \, {\cal O}_{{\bf p}_1} (\eta_1)
  \, ... \, {\cal O}_{{\bf
      p}_N} (\eta_N) \, \rangle^{c'} \over P_\Delta ({\rm p}, \eta)} = 0 \, ,}
\end{eqnarray}
where the power spectrum of displacement is defined by
$\langle \Delta^i_{\bf p} \Delta^j_{{\bf p}'} {}^* \rangle = 
(2\pi)^3 \delta_D ({\bf p} - {\bf p}')  ({\rm p}^i {\rm p}^j / {\rm
  p}^2) P_\Delta ({\rm p})$.
It is important to emphasize that the Eulerian space
consistency relation (Eq.\ \ref{KRPPEuler}) 
already yields a vanishing right hand side
if $\eta_1 = \eta_2 ... = \eta_N$. The Lagrangian space
consistency relation (Eq.\ \ref{KRPPLag00} or \ref{KRPPLag}),
on the other hand, has a vanishing right hand side
{\it regardless} of what the times $\eta_1, ... , \eta_N$ 
happen to be.
The consistency relation can also
be viewed as a statement about how the
squeezed correlation function (normalized by the
soft power spectrum) scales with the soft momentum:
the Eulerian space consistency relation states that
such a squeezed correlation function goes like
${\rm k}^0$ (${\bf k}$ is the soft
momentum); the Lagrangian space consistency relation
states that there is no ${\rm p}^0$ term,
and at best there is a ${\rm
  p}^\epsilon$ contribution
with $\epsilon > 0$.

The simplest way to derive Eq.\ (\ref{KRPPLag}) is to work out the implications
of the KRPP symmetry entirely within Lagrangian space. This is done in
\S \ref{WardLag}. We perform a perturbative 
check of this Lagrangian space consistency relation using Lagrangian
perturbation theory in \S \ref{PTcheck}. Because the Eulerian space
and the Lagrangian space
relations look so different, as a further check, we show how one can be obtained from 
the other 
in \S \ref{nonPTcheck}. 
Since observations are performed in Eulerian, not Lagrangian, space, the fact that the consistency relation takes a particularly simple
form in Lagrangian space is mainly of theoretical interest.
The simplicity of the Lagrangian space consistency relation should
not be interpreted as the lack of physical content, however -- in the Lagrangian as well as in the Eulerian picture,
the consistency relation can be viewed as a test of the single-field
initial condition and of the equivalence principle. Rather, the simplicity
suggests that an analytical understanding of nonlinear clustering
might be most promising in Lagrangian space. This is
discussed in \S \ref{discuss}.

\section{The Lagrangian space consistency relation:\ derivation}%
\label{WardLag}

After a brief review of notation, we derive our main result --
the Lagrangian space consistency relation -- using the background wave argument phrased entirely in Lagrangian space.

\subsection{Notation}
\label{notation}

We use $\textbf{q}$ to denote the Lagrangian space coordinate of a particle,
which coincides with its initial position, and $\textbf{x}$ to denote the Eulerian
space coordinate which is its position at a later time.
To be definite, in cases where multiple components are present,
the Lagrangian space coordinate $\textbf{q}$ refers to that of the dark
matter particle, which has only gravitational interactions.\footnote{Our derivation of the Lagrangian space consistency relation 
would go through even if we chose the Lagrangian coordinate to track
other constituents of the universe.}
Both coordinates are defined in comoving space where the expansion of the universe
is scaled out. The (dark matter) displacement $\pmb{\Delta}$ is the difference:
\begin{equation}
  \textbf{x}\left( \textbf{q},\eta \right)=\textbf{q}+\pmb{\Delta}\left( \textbf{q},\eta \right) \, .
  \label{displaceDelta}
\end{equation}
The (dark matter) velocity is given by the conformal time derivative of  $\pmb{\Delta}$ at a fixed
Lagrangian coordinate:
\begin{equation}
\label{velocity}
  \pmb{v}\left( \textbf{q},\eta \right)={\partial \pmb{\Delta}
\over \partial \eta} \Big|_{\bf q}\, .
\end{equation}
The (dark matter) overdensity $\delta$ can be obtained by mass conservation, assuming
the initial overdensity is negligible:
\begin{eqnarray}
\label{deltaJ}
1 + \delta ({\bf x}, \eta) =| J({\bf q}, \eta) |^{-1}
\end{eqnarray}
with $J\left( \textbf{q},\eta \right)$ being the Jacobian relating the volume elements in Eulerian and Lagrangian space:
\begin{equation}
\label{Jacobian}
J({\bf q}, \eta) \equiv {\,\det} \left[ {\partial x^i({\bf q},
    \eta) \over \partial q^j} \right] \, .
\end{equation}
The Jacobian $J$ as a function of $\bf q$ is well-defined even in the presence
of multiple-streaming -- where a single ${\bf x}$ corresponds to multiple ${\bf q}$'s --
but Eq.\ (\ref{deltaJ}) requires modification in that case:
\begin{equation}
\label{deltaJacobian}
1 + \delta\left( \textbf{x},\eta \right)= \sum_{\textbf{x} = \textbf{q} + \pmb{\Delta}(\textbf{q}, \eta)} 
|J\left( \textbf{q},\eta \right) |^{-1} \, ,
\end{equation}
where the sum is over all $\textbf{q}$'s that reach the same
$\textbf{x}$.

Suppose we have some LSS observable ${\cal O}$. This could represent
many different quantities, such as mass overdensity or galaxy number
overdensity.\footnote{\label{densitydef} Unless otherwise stated, whenever we discuss mass or galaxy
  density, we mean the mass or galaxy count per unit {\it Eulerian}
  space volume. Such a quantity can of course be expressed as a function of
either Eulerian space coordinate ${\bf x}$ or Lagrangian space
coordinate ${\bf q}$.
}
What we typically observe is ${\cal O}$ as a
function of ${\bf x}$ (and possibly time, which we suppress). Given this function ${\cal
  O}({\bf x})$, one can define unambiguously a corresponding function
of ${\bf q}$:\ ${\cal O}({\bf x}({\bf q}))$. In other words, suppose we
are interested in the value of ${\cal O}$ at a Lagrangian location 
${\bf q}$:\ we can define it by working out the ${\bf x}$ that ${\bf q}$
maps to, and then evaluating ${\cal O}({\bf x})$. This
procedure is well defined even if multiple ${\bf q}$'s map to the same
${\bf x}$, which is expected to happen for dark matter in the
nonlinear regime.

Some quantities defined in Lagrangian space, on the other hand, might not
have an unambiguous meaning in Eulerian space.
For instance, the velocity $\pmb{v}$ given in Eq.\ (\ref{velocity})
is defined for a dark matter particle labeled by 
the Lagrangian coordinate $\textbf{q}$. 
At an Eulerian position $\textbf{x}$ where multiple
Lagrangian streams cross, additional inputs are required to
define a velocity;\ a reasonable definition is:
\begin{eqnarray}
{\rm average\,\,} \pmb{v} = \frac{1}{\mathcal{N}}\sum_{\textbf{x} = 
\textbf{q} + \pmb{\Delta}} \pmb{v} (\textbf{q}) \, ,
\end{eqnarray}
where the sum is over all $\textbf{q}$'s that map to the same
$\textbf{x}$, and $\mathcal{N}$ is the number of such $\textbf{q}$'s.  This gives a mass weighted velocity.

It is interesting to contrast the Fourier transform in Lagrangian versus
Eulerian space, as described by Eq.\ (\ref{Fourier}). In particular,
the Eulerian space Fourier transform can be rewritten as (suppressing
time dependence):
\begin{eqnarray}
\label{Ok}
{\cal O}_{\bf k} = \int d^3 {\bf x} \, {\cal O} ({\bf x}) e^{i {\bf k} \cdot
  {\bf x}}  = \int d^3 {\bf q} \,  J({\bf q}) \, {\cal O}({\bf x}({\bf q})) e^{i
  {\bf k} \cdot ({\bf q} + \pmb{\Delta})} \, ,
\end{eqnarray}
where $J$ comes without absolute value;
this expression remains valid in the presence of
multiple streaming. 
Note how an Eulerian space Fourier transform
of ${\cal O}({\bf x})$
can be interpreted as a Lagrangian space Fourier
transform of $J({\bf q}) {\cal O}({\bf x}({\bf q})) e^{i
{\bf k} \cdot {\pmb{\Delta}}}$.

\subsection{Derivation from the displacement symmetry}
\label{derivation}


We now deduce our main result, making use of a master formula derived
in an earlier paper \cite{us}. At the heart of the consistency relation is the
existence of a nonlinearly realized symmetry, under which some field
-- the Goldstone boson or pion $\pi$ -- transforms as
$\pi \to \pi + \Delta_{\rm lin.} \pi + \Delta_{\rm nl.}\pi$.
Here, $\Delta_{\rm lin.} \pi$ is the part of the transformation
that is linear in $\pi$, and $\Delta_{\rm nl.}\pi$ is the part of the
transformation that is independent of $\pi$ (i.e., nonlinear in $\pi$,
though `sub-linear' or `inhomogeneous' would be a better description).
The fact that $\Delta_{\rm nl.}\pi \ne 0$ is the sign of a nonlinearly
realized, or spontaneously broken, symmetry.
At the same time, there are other fields or observables ${\cal O}$
that could have their own linear and/or nonlinear transformations.
The master formula (in momentum space) reads \cite{us}:
\begin{eqnarray}
\label{consistencymasterrelation}
  \int \frac{d^3 \textbf{p}}{(2\pi)^3} \frac{\langle \pi_{\textbf{p}}
    \mathcal{O}_{\textbf{p}_1} \cdots
    \mathcal{O}_{\textbf{p}_N}\rangle^c}{P_{\pi}({\rm p})}\Delta_{\rm
    nl.}\pi^{*}_{\textbf{p}} = \Delta_{\rm
    lin.}\langle\mathcal{O}_{\textbf{p}_1} \cdots
  \mathcal{O}_{\textbf{p}_N}\rangle^c \, ,
\end{eqnarray}
where $\langle ... \rangle^c$ refers to the connected correlation
function {\it without} removing the overall delta function
(as opposed to $\langle ... \rangle^{c'}$ which has the delta function
removed). Note how it is the nonlinear transformation
of $\pi$ and the linear transformation of ${\cal O}$
that show up on the left and the right respectively.
Note also that the ${\cal O}$'s need not even be the same
observable. Nor do $\pi$ and the ${\cal O}$'s need be at the same
time: they can be at {\it arbitrary, potentially different, times}.
The derivation of this master formula made no
assumption about whether the quantities (or the
Fourier transform thereof) are defined in Eulerian or
Lagrangian space.
We are thus free to use it in either.
This master relation can be used to derive the large scale structure analog of
Ward identities or soft-pion theorems in particle physics.

As a warm-up, let us first apply this formula to
a simple system that involves the dark matter only. 
The dynamics is described by:\ (1) ${\bf x} = {\bf q} + \pmb{\Delta}$
as in 
Eq.\ (\ref{displaceDelta});\ (2) the dark mater overdensity $\delta$ determined
by the Jacobian as in Eq.\ (\ref{deltaJacobian});\ (3) the displacement
$\pmb{\Delta}$ which evolves according to: 
\begin{eqnarray}
\label{eomDisplacement}
{\partial^2 \pmb{\Delta} \over \partial\eta^2}  \Big|_{\bf q}
+ {a' \over a} {\partial \pmb{\Delta} \over \partial\eta} \Big|_{\bf q} = -\pmb{\nabla}_x \Phi \, ,
\end{eqnarray}
where $a$ is the scale factor, $a'$ is its derivative with respect to
conformal time $\eta$, $\Phi$ is the gravitational potential and
$\pmb{\nabla}_x$ is the partial derivative with respect to ${\bf x}$;
lastly (4) the Poisson equation:
\begin{eqnarray}
\label{Poisson}
\nabla^2_x \Phi = 4 \pi G a^2 \bar\rho \delta \, ,
\end{eqnarray}
where $G$ is the Newton constant and $\bar\rho$ is the mean mass density.

This system has the following symmetry:
\begin{eqnarray}
\label{symmetryLag}
\boxed{
{\bf q} \rightarrow {\bf q} \quad , \quad
\pmb{\Delta} \rightarrow \pmb{\Delta} + {\bf n}(\eta) \quad , \quad
\Phi \rightarrow \Phi - \left( {\bf n}'' + {a' \over a} {\bf n}'
\right) \cdot {\bf x} \, ,}
\end{eqnarray}
where ${\bf n}(\eta)$ is a function of time alone.
We will refer to this as the {\it displacement symmetry}. 
Note how $\pmb{\Delta}$ shifts by a nonlinear (or sub-linear) amount
and can be thought of as our Goldstone boson.
The same is true for $\Phi$.
The interesting point is that the mass overdensity
$\delta$ does not transform at all under this symmetry.
Nor are ${\bf q}$ or $\eta$ transformed.
Applying the master formula, choosing the observable ${\cal O} =
\delta$, we thus find:
\begin{eqnarray}
\label{KRPPLag0}
\lim_{{\bf p} \rightarrow 0} {\langle \pmb{\Delta}_{\bf p} \delta_{\bf p_1} ... \delta_{\bf p_N}
\rangle^{c'} \over P_\Delta ({\rm p})} = 0 \, .
\end{eqnarray}
Here, we have used the fact that the nonlinear transformation of
$\pmb{\Delta}$ in Fourier space is
$\Delta_{\rm nl.} \pmb{\Delta}_{\bf p} = {\bf n}(\eta) (2\pi)^3 \delta_D
({\bf p})$, where $\delta_D ({\bf p})$ is the Dirac delta function.
We have also removed the overall momentum-conserving delta function.
The power spectrum of displacement $P_\Delta$ is as defined in \S
\ref{intro}.

Two comments are in order before we proceed to generalize this
derivation to more realistic, astrophysically complex observables.
First, while the Lagrangian coordinate ${\bf q}$ does not transform
under the symmetry of interest, 
the Eulerian coordinate ${\bf x} = {\bf q} + \pmb{\Delta}$ does,
because the displacement $\pmb{\Delta}$ shifts. This implies that an observable
like $\delta$, when expressed as a function of ${\bf x}$, transforms as:
$\delta \rightarrow \delta + \Delta_{\rm lin.} \delta$ with
$\Delta_{\rm lin.} \delta = \pmb{\Delta} \cdot \pmb{\nabla} \delta$. 
Plugging this into the master formula
Eq.\ (\ref{consistencymasterrelation}), we see that there is a
non-vanishing right hand side, unlike the situation in Lagrangian
space where $\delta$ expressed as a function of $\bf{q}$ does not
shift at all. The is the fundamental reason why the KRPP consistency
relation takes a more complicated form in Eulerian space
(Eq.\ \ref{KRPPEuler}) than in Lagrangian space (Eq.\ \ref{KRPPLag0}).

Second, the reader might wonder about the validity of
our application of the master formula:\
on the one hand, the master relation is
phrased in terms of a scalar pion; on the other,
our application effectively uses the vector displacement
$\pmb{\Delta}$ as the pion.
The short answer is that the master formula is applicable
to any field $\pi$ that shifts nonlinearly under the symmetry
of interest; one can use it for each component of $\pmb{\Delta}$
for instance. The long answer is:\ since $\pmb{\Delta}$ is used
in the consistency relation only as a soft (long wavelength) mode,
one is justified in treating it as a gradient mode (assuming growing
mode initial condition) with $\pmb{\Delta} = \pmb{\nabla}_q \pi$
with $\pi$ playing the role of the displacement potential. 
The master formula can then be applied with the displacement
potential as the pion. The resulting consistency relation can be
shown to be equivalent to the one we have derived.\footnote{
There are actually two different nonlinear
realized symmetries associated with the displacement potential. One is shifting it
by a constant or a function of time (but not space). The other is shifting
it by a linear gradient, i.e., $\pi \rightarrow \pi + {\bf n} \cdot {\bf
  q}$ where ${\bf n}$ is the same as that in Eq.\ (\ref{symmetryLag}). 
There are as a result two consistency relations which can
be succinctly combined into one, Eq.\ (\ref{KRPPLag0}). 
See our earlier paper \cite{us} for further discussions.}

Let us turn to the derivation of a stronger form of the Lagrangian
space consistency relation.
So far, we have focused on a simple system of dark matter particles
that interact only gravitationally, as embodied in
Eqs.\ (\ref{eomDisplacement}) and (\ref{Poisson}). 
Let us consider the addition of galaxies into the mix.
They have their own number overdensity $\delta_g$, 
displacement $\pmb{\Delta}_g$ and velocity ${\bf v}_g =
\pmb{\Delta}'_g$. Their number density is not necessarily conserved
by evolution, since galaxies can form and merge:
\begin{eqnarray}
\label{R}
\delta_g' + (1+ \delta_g) \pmb{\nabla}_x \cdot {\bf v}_g = R_g \, ,
\end{eqnarray}
where ${}'$ refers to conformal time derivative at a fixed Lagrangian
coordinate and $R_g$ is a source term that incorporates the formation and
merger rates. The equation of motion for the galaxies is:
\begin{eqnarray}
\label{F}
\pmb{\Delta}''_g + {a' \over a} \pmb{\Delta}'_g = - \pmb{\nabla}_x \Phi 
+ {\bf F}_g \, ,
\end{eqnarray}
where ${\bf F}_g$ encodes additional forces that might act on galaxies, such as gas pressure, dynamical friction et cetera.
The gravitational potential $\Phi$ is determined of course by the
Poisson equation (\ref{Poisson}) as before.

The displacement symmetry of Eq.\ (\ref{symmetryLag}) can be extended to include
also:
\begin{eqnarray}
\label{symmetryLag2}
\pmb{\Delta}_g \rightarrow \pmb{\Delta}_g + {\bf n}(\eta) \, ,
\end{eqnarray}
which also implies ${\bf v}_g \rightarrow {\bf v}_g + {\bf n}'$. 
The galaxy overdensity $\delta_g$, like its dark matter counterpart,
does not transform under this symmetry. Eqs.\ (\ref{symmetryLag}) and (\ref{symmetryLag2})
represent the displacement symmetry of the combined dark-matter-galaxies system,
as long as $R_g$ and ${\bf F}_g$ depend only on (dark matter/galaxy)
densities and gradients
of (dark matter/galaxy) velocities -- recall that neither shifts under our
symmetry. 
What happens if $R_g$ and/or ${\bf F}_g$ depends on velocities as
opposed to gradients of velocities? In that case, shifting velocities
by a spatially constant amount would affect the galaxy formation and
dynamics -- this is a violation of the equivalence principle which
states that local physical processes (such as galaxy formation, mergers
and motion) should not be dependent on the absolute state of motion.
Note that a dependence on the dark-matter-galaxy velocity {\it difference} ${\bf
  v} - {\bf v}_g$, on the other hand, is consistent with the equivalence principle, and
the velocity difference is indeed unchanged under our symmetry.
Thus, as long as the equivalence principle is respected, whether
$R_g$ and ${\bf F}_g$ depend on densities, gradients of velocities or
velocity differences, the displacement symmetry holds.
Furthermore, the same statement is expected to be valid in
a system with many different species, such as baryons, galaxies or
even dark matter of different kinds.
The argument that leads to Eq.\ (\ref{KRPPLag0}) can be rerun to
give the more general Lagrangian space consistency relation:
\begin{eqnarray}
\label{KRPPLag1}
\boxed{
\lim_{{\bf p} \rightarrow 0}
{\langle \pmb{\Delta}_{\bf p} (\eta) \, {\cal O}_{{\bf p}_1} (\eta_1)
  \, ... \, {\cal O}_{{\bf
      p}_N} (\eta_N) \, \rangle^{c'} \over P_\Delta ({\rm p}, \eta)} = 0 \, ,}
\end{eqnarray}
where ${\cal O}$ is any observable that has no linear shift
under the displacement symmetry --
this includes for instance
the densities, displacements and velocities of
the galaxies and of dark matter.\footnote{\label{linearshift} The reader might wonder: given that 
the Lagrangian coordinate ${\bf q}$ does not get transformed at all
under the displacement symmetry, is there any observable that
has a linear shift? The answer is yes. For instance, the
combination ${\cal O} = {\bf v} \delta$ transforms to
$({\bf v} + {\bf n}') \delta$ giving a shift that is linear
in the fluctuation variable $\delta$. 
}
Note that the ${\cal O}$'s need not
be the same observables. We have restored the explicit
time-dependence of each fluctuation variable to emphasize
the fact that the times need not be equal.
Note also that we have chosen the dark matter displacement to be the pion.
We could have chosen the galaxy displacement instead.
Assuming that gravity is the dominant interaction on large scales
and adiabatic initial conditions, the two displacements are expected to
coincide in any case in the soft limit.
Furthermore, we could have chosen the velocity instead of the
displacement as the soft-pion, in which case Eq.\ (\ref{KRPPLag00})
follows.

\section{The Lagrangian space consistency relation:\ checks}
\label{checks}

The above derivation of the Lagrangian space consistency relation 
is a bit terse, and the form the relation takes is surprisingly
simple. It is thus worth performing some non-trivial checks of the
relation. We will first do this using second order Lagrangian
perturbation theory (\S \ref{PTcheck}).
Then, in \S \ref{nonPTcheck}, we demonstrate how the Eulerian space
consistency relation can be derived from its counterpart in Lagrangian
space.

\subsection{Perturbative check}
\label{PTcheck}

Let us perform an explicit check of  Eq.\ (\ref{KRPPLag1}) using
second-order Lagrangian space perturbation theory.  For simplicity, we will
focus on the case where the only species present is
dark matter and the observable ${\cal O} = \delta$. We will 
confine the discussion to the squeezed three-point function; extension to a general
(N+1)-point function is straightforward.
Expanding Eq.\ (\ref{deltaJ}) to second order, we have
\begin{equation}
\delta(\textbf{x}(\textbf{q}, \eta), \eta) = -\nabla_\textbf{q} \cdot
\pmb{\Delta} + \frac{1}{2}(\nabla_\textbf{q} \cdot \pmb{\Delta})^2 +
\frac{1}{2}\nabla_{\textbf{q}^i}\pmb{\Delta}^j \nabla_{\textbf{q}^j}
\pmb{\Delta}^i \, .
\end{equation}
Expanding out $\delta_{\textbf{p}} = \delta^{(1)}_{\textbf{p}} +
\delta^{(2)}_{\textbf{p}} + \cdots$, $\pmb{\Delta}_{\textbf{p}} =
\pmb{\Delta}^{(1)}_{\textbf{p}} + \pmb{\Delta}^{(2)}_{\textbf{p}} +
\cdots$, and plugging into
Eqs.\ (\ref{eomDisplacement}) and (\ref{Poisson}), we have
\cite{Bernardeau:2001qr}:
\begin{equation}
\begin{split}
\Delta^j_{\textbf{p}} {}^{(1)} (\eta) &=  {-i \, {\rm p}^j \over {\rm
    p}^2} \,
\delta^{(1)}_{\textbf{p}} (\eta) \, , \\
\Delta^j_{\textbf{p}} {}^{(2)} (\eta) &= {1\over 2} {D_2 (\eta) \over D(\eta)^2}
{i \,{\rm p}^j \over {\rm p}^2}
\int {d^3 {\bf p}_A d^3 {\bf p}_B \over (2\pi)^3}    
\delta_D ({\textbf{p}_A} + {\textbf{p}_B} - {\textbf{p}}) 
\, \left( 1 - { ( {\textbf{p}_A} \cdot {\textbf{p}_B} )^2
      \over {\rm p}_A^2 {\rm p}_B^2} \right) 
\, \delta^{(1)}_{\textbf{p}_A} (\eta) \delta^{(1)}_{\textbf{p}_B} (\eta) \, ,
\\
\delta^{(2)}_{\textbf{p}} (\eta) &= {1\over 2} \int \frac{d^3
  {\bf p}_A d^3 {\bf p}_B }{(2\pi)^3}
\delta_D (\textbf{p}_A + \textbf{p}_B - \textbf{p}) 
\left(
1 - {D_2 (\eta) \over D(\eta)^2} 
+ 
{ ({\textbf{p}}_A
  \cdot {\textbf{p}}_B )^2 \over {\rm p}_A^2 {\rm p}_B^2}
\left[ 1 + {D_2 (\eta) \over D(\eta)^2} \right] 
\right) \,
\delta^{(1)}_{\textbf{p}_A} (\eta) \delta^{(1)}_{\textbf{p}_B} (\eta)
\, ,
\\
\end{split}
\end{equation}
where $D$ is the linear growth factor determining the
time-dependence of the first order displacement (and density), and $D_2$ is the 
second order growth factor determining that of the second order displacement.
They satisfy the equations:
\begin{eqnarray}
\label{D2}
&& D'' + {a' \over a} D' - 4\pi G a^2 \bar\rho D = 0 \, , \nonumber \\
&& D_2'' + {a' \over a} D_2' - 4\pi G a^2 \bar\rho D_2 = -4\pi G a^2 \bar\rho
D^2 \, .
\end{eqnarray}
For instance, in a flat universe with $\Omega_m = 1$, $D_2 = -3 D^2/7$. 
Using these expressions, we can work out the 
lowest order contributions to the relevant squeezed
bispectrum:
\begin{equation}
\label{PTbispectrum}
\begin{split}
\langle \Delta^j_{\bf p} (\eta) \delta_{{\bf p}_1} (\eta_1)
\delta_{{\bf p}_2} (\eta_2) \rangle
& = \langle \Delta^j_{\bf p} {}^{(2)}(\eta) \delta_{{\bf p}_1}^{(1)} (\eta_1)
\delta_{{\bf p}_2}^{(1)} (\eta_2) \rangle
+ \langle \Delta^j_{\bf p} {}^{(1)} (\eta) \delta_{{\bf p}_1}^{(2)} (\eta_1)
\delta_{{\bf p}_2}^{(1)} (\eta_2) \rangle
+ \langle \Delta^j_{\bf p} {}^{(1)} (\eta) \delta_{{\bf p}_1}^{(1)} (\eta_1)
\delta_{{\bf p}_2}^{(2)} (\eta_2) \rangle \\
& = O({\rm p}^j) + O({\rm p}^j P_\Delta ({\rm p})) \, ,
\end{split}
\end{equation}
where we spell out the dependence on the soft momentum
${\bf p}$:\ the $O({\rm p}^j)$ piece comes from
the first term on the right in the first line, and
the $O({\rm p}^j P_\Delta ({\rm p}))$ piece comes from the other
two terms. We have used the fact that
$1 - [({\bf p}_A \cdot {\bf p}_B)^2/{\rm p}_A^2 {\rm p}_B^2]
= O({\rm p}^2)$ for ${\bf p}_A + {\bf p}_B = {\bf p}$. 
The $O({\rm p}^j P_\Delta ({\rm p}))$
piece is obviously compatible with the Lagrangian space consistency relation;
i.e., it gives $\langle \Delta^j_{\bf p} \delta_{{\bf p}_1}
\delta_{{\bf p}_2} \rangle^{c'} / P_\Delta ({\rm p}) = 0$ in the ${\bf p}
\rightarrow 0$ limit. The $O({\rm p}^j)$ piece does the same,
provided the power spectrum $P_\Delta ({\rm p})$ is not too blue.
Parameterizing the power spectrum $P_\Delta ({\rm p}) \propto {\rm p}^{n-2}$ in
the low momentum limit, the consistency relation holds as long
as $n < 3$. Exactly the same condition is needed for the Eulerian space
consistency relation (see e.g.\ \cite{us}).\footnote{The Eulerian space consistency relation is often given in
a form where $\delta$ is used as the soft mode. One might be
tempted to do the same for the Lagrangian space consistency relation.
However, one can check using perturbation theory that such a consistency
relation would have required $n < 1$, a condition
considerably stronger than expected. This is related to
the fact that $\delta^{(2)}_{\bf p}$ does not vanish in the ${\bf p}
\rightarrow 0$ limit, unlike $\Delta^j_{\bf p} {}^{(2)}$.
}

\subsection{Recovering the Eulerian space consistency relation
from Lagrangian space}%
\label{nonPTcheck}

The consistency relation takes such a different form in Lagrangian
versus Eulerian space that it is worth considering how one
can be derived from the other.
Let us compute the following:
\begin{equation}
\label{Edef}
\begin{split}
E & \equiv E_L + E_R \\
E_L & \equiv \lim_{{\bf k} \rightarrow 0}
{\langle v^j_{\bf k} (\eta) \, {\delta}_{{\bf k}_1} (\eta_1)  \,  {\delta}_{{\bf k}_2} (\eta_2) \, \rangle^{c'} \over
P_v ({\rm k}, \eta)} \quad , \quad E_R \equiv -
{i {\rm k}^j} \sum_{a=1}^2 {D(\eta_a) \over D'(\eta)} 
{{\bf k} \cdot {{\bf k}_a} \over {\rm k}^2} \langle {\delta}_{{\bf
    k}_1}  (\eta_1) \, {\delta}_{{\bf k}_2} (\eta_2) \,
\rangle^{c'} \, .
\end{split}
\end{equation}
The Eulerian space consistency condition is the statement that $E =
0$. We will content ourselves with deriving this -- a special
case of the more general Eulerian space consistency relation
(\ref{KRPPEuler}) -- from the Lagrangian space consistency relation.

To relate $E$ to quantities in Lagrangian space, we will
slightly abuse our notation. So far, we have been using ${\bf k}$ for
the Eulerian space momentum and ${\bf p}$ for the Lagrangian space
momentum. For instance, $\delta_{{\bf k}_1}$ is defined as
\begin{eqnarray}
\delta_{{\bf k}_1} = \int d^3 {\bf x} \, \delta ({\bf x}) \, e^{i {\bf k}_1
  \cdot {\bf x}} \, .
\end{eqnarray}
Let us rewrite this as
\begin{equation}
\label{deltakrewrite}
\begin{split}
\delta_{{\bf k}_1 = {\bf p}_1} = & \int d^3 {\bf q} \, J({\bf q})\, \delta
({\bf x}({\bf q})) \, e^{i {\bf p}_1
  \cdot ({\bf q} + \pmb{\Delta}({\bf q}))}  \\
= & \int d^3 {\bf q} \, J({\bf q})\, \delta
({\bf x}({\bf q})) \, e^{i {\bf p}_1
  \cdot {\bf q}}  + i {\rm p}_1^m
\int d^3 {\bf q} \, J({\bf q})\, \delta
({\bf x}({\bf q})) \, \Delta^m ({\bf q}) \, e^{i {\bf p}_1
  \cdot {\bf q}}  \\
& - {1\over 2} {\rm p}_1^m {\rm p}_1^n
\int d^3 {\bf q} \, J({\bf q})\, \delta
({\bf x}({\bf q})) \, \Delta^m ({\bf q}) \Delta^n ({\bf q}) \, e^{i {\bf p}_1
  \cdot {\bf q}}  + ...
\, ,
\end{split}
\end{equation}
where $J({\bf q})$ is
defined by Eq.\ (\ref{Jacobian}) (with no absolute value).
Defining
\begin{eqnarray}
\tilde\delta({\bf q}) \equiv J({\bf q}) \delta({\bf x}({\bf q})) \, ,
\end{eqnarray}
we see that the first term on the right of Eq.\ (\ref{deltakrewrite})
is $\tilde\delta_{{\bf p}_1}$, i.e.\ the Fourier transform
of $\tilde\delta$ in {\it Lagrangian space}.
The other terms on the right can likewise be thought of
as the Fourier transform of some quantity in Lagrangian space.
This is why we introduce 
${\bf p}_1$ as the momentum label for these Fourier components.
On the other hand, upon summation, they give the quantity on
the left $\delta_{{\bf k}_1 = {\bf p}_1}$ which is the
Fourier transform of density in {\it Eulerian space} -- this is
why we use ${\bf k}_1$ as its momentum label; it just happens
to take on the numerical value ${\bf p}_1$ which conveniently
gives us the appropriate momentum label for quantities on the right.
It is worth emphasizing that our definitions are general, in that they
are valid even in the presence of multiple-streaming (see \S \ref{notation}). 

The expansion in terms of $\Delta$ in Eq.\ (\ref{deltakrewrite}) is
purely formal. In the nonlinear regime, there is no sense in which
$\Delta$ is small. The expansion provides a convenient way to
relate the Fourier transform in Eulerian space to the Fourier transform
in Lagrangian space. We will argue $E = 0$ holds to arbitrary order
in a power series expansion.

For the soft mode, we have
\begin{eqnarray}
\label{vvkp}
v^j_{{\bf k}={\bf p}} = v^j_{{\bf p}} + ... \, ,
\end{eqnarray}
This is where our abuse of notation is the most egregious:
on the left is the velocity Fourier transformed in Eulerian space;
on the right is the velocity Fourier transformed in Lagrangian space.
They agree only to lowest order in perturbations. For the soft-mode,
ignoring the higher order corrections is permissible:\ the higher order
corrections will give higher powers of the soft momentum ${\bf p}$
compared to what is kept in the consistency relation, provided
that the soft power spectrum 
$P_v ({\rm p})$ or $P_\Delta ({\rm p})$ is not too blue (see \S
\ref{PTcheck}).
Similarly, it can be shown that in the soft limit, there is no need
to distinguish between $P_v$ in Lagrangian versus
Eulerian space.\footnote{
It is also worth emphasizing that the notion of a well-defined
velocity in Eulerian space is valid only when
multiple-streaming is ignored. This is acceptable for the soft-mode.
We do not assume single-streaming for the hard modes.
}

Let us substitute Eq.\ (\ref{deltakrewrite}) for the hard modes, and
Eq.\ (\ref{vvkp}) for the soft mode, into the expression for $E$ in
Eq.\ (\ref{Edef}).
Consider first what contributes to $E_L$:
\begin{equation}
\label{vdd}
\begin{split}
\langle v^j_{{\bf k} = {\bf p}} \delta_{{\bf k}_1 = {\bf p}_1}
\delta_{{\bf k}_2 = {\bf p}_2} \rangle
= & \langle v^j_{\bf p} \tilde\delta_{{\bf p}_1} \tilde\delta_{{\bf
    p}_2} \rangle \\
& + \Bigg[ i {\rm p}_1^m \int {d^3 {\bf p}_A \over (2\pi)^3} \langle v^j_{\bf p}
\Delta^m_{{\bf p}_A} (\eta_1) \tilde\delta_{{\bf p}_1 - {\bf p}_A} \tilde\delta_{{\bf
    p}_2} \rangle + 1 \leftrightarrow 2 \Bigg] \\
& - \Bigg[ {\rm p}_1^m {\rm p}_2^n
\int  {d^3 {\bf p}_A \over (2\pi)^3}  {d^3 {\bf p}_B \over (2\pi)^3}
\langle v^j_{\bf p} \tilde\delta_{{\bf p}_1 - {\bf p}_A}
\Delta^m_{{\bf p}_A} (\eta_1) \tilde\delta_{{\bf p}_2-{\bf p}_B}
\Delta^n_{{\bf p}_B} (\eta_2) \rangle \\
& \quad + \Big[ {1\over 2} 
{\rm p}_1^m {\rm p}_1^n
\int  {d^3 {\bf p}_A \over (2\pi)^3}  {d^3 {\bf p}_B \over (2\pi)^3}
\langle v^j_{\bf p} \tilde\delta_{{\bf p}_1 - {\bf p}_A-{\bf p}_B}
\Delta^m_{{\bf p}_A} (\eta_1) 
\Delta^n_{{\bf p}_B} (\eta_1)
\tilde\delta_{{\bf p}_2}
\rangle + (1 \leftrightarrow 2)
\Big] \Bigg]\\
& + O(\Delta^3) + ...
\end{split}
\end{equation}
where we have largely suppressed the time-dependence to
minimize clutter ($\eta$ for the soft mode, and $\eta_1$ and $\eta_2$
respectively for the hard modes), except
for variables with internal momenta. 
We emphasize that the expansion in $\Delta$ is purely formal,
and comes entirely from expanding
$e^{i {\bf p}_1  \cdot \pmb{\Delta}}$ or $e^{i {\bf p_2} \cdot
  \pmb{\Delta}}$. 
The first term on the right can be set to zero by virtue of the
Lagrangian space consistency condition
(keeping in mind that this term is divided by $P_v ({\rm p})$ as part of
the quantity $E_L$).
We will be assuming the Lagrangian space consistency
relation in its general form (Eq.\ \ref{KRPPLag00}):
\begin{equation}
\label{KRPPLag00repeat}
\lim_{{\bf p} \rightarrow 0}
{\langle \pmb{v}_{\bf p} (\eta) \,  {\cal O}_{{\bf p}_1} (\eta_1) \,
  ... \, {\cal O}_{{\bf
      p}_N} (\eta_N) \, \rangle^{c'} \over
P_v ({\rm p}, \eta)} = 0 \, ,
\end{equation}
where the observables at hard momenta need not be the same observable.  Finally, note that while we are interested in the connected part of the correlator on the left hand side of Eq.\ (\ref{vdd}), the correlators on the right hand side are the full correlators, minus the contributions where some proper subset of the original hard and soft momenta sum to zero.  In particular, the correlators on the right hand side of Eq.\ (\ref{vdd}) contain both connected and disconnected pieces.

The second term on the right of Eq.\ (\ref{vdd}), formally $O(\Delta)$, equals
\begin{equation}
\label{LHSDelta1}
\begin{split}
 i {\rm p}_1^m \int {d^3 {\bf p}_A \over (2\pi)^3} 
& \Bigg[ \langle v^j_{\bf p} \Delta^m_{{\bf p}_A} (\eta_1) \rangle \langle
\tilde\delta_{{\bf p}_1 - {\bf p}_A} \tilde\delta_{{\bf p}_2} \rangle \\
& + \langle v^j_{\bf p} \tilde\delta_{{\bf p}_1 - {\bf p}_A} \rangle \langle
\Delta^m_{{\bf p}_A} (\eta_1) \tilde\delta_{{\bf p}_2} \rangle 
+ \langle v^j_{\bf p} \tilde\delta_{{\bf p}_2} \rangle \langle
\Delta^m_{{\bf p}_A} (\eta_1) \tilde\delta_{{\bf p}_1 - {\bf p}_A}
\rangle \\
& + \langle v^j_{\bf p} \Delta^m_{{\bf p}_A} (\eta_1)
\tilde\delta_{{\bf p}_1 - {\bf p}_A} \tilde\delta_{{\bf p}_2}
\rangle^c \Bigg] + (1 \leftrightarrow 2) \, ,\\
\end{split}
\end{equation}
where the connected trispectrum term $\langle v^j_{\rm p} ... \rangle^c$ (anticipating
division by $P_v ({\rm p})$) can be
set to zero using the Lagrangian space consistency relation,
the terms involving $\langle v^j_{\bf p} \tilde\delta_{{\bf p}_1 -
  {\bf p}_A} \rangle$, $\langle v^j_{\bf p} \tilde\delta_{{\bf p}_2} \rangle$
and the like have one more power of the soft momentum ${\bf p}$ (and are thus subdominant)
compared to terms involving $\langle v^j_{\rm p} \Delta^m_{{\bf p}_A}
\rangle$ which give:
\begin{equation}
\label{LHSDelta1result}
\begin{split}
(2\pi)^3 \delta_D ({\bf p}_1 + {\bf p_2} + {\bf p}) \, P_v ({\rm p},
\eta) \, i{\rm p}^j \,
{D(\eta_1) \over D'(\eta)} {{\bf p}_1 \cdot {\bf p} \over {\rm p}^2}
\langle \tilde \delta_{{\bf p}_1 + {\bf p}} (\eta_1) \tilde
\delta_{{\bf p}_2} (\eta_2)
\rangle^{c'} \, + (1 \leftrightarrow 2) \, .
\end{split}
\end{equation}

The third term on the right of Eq.\ (\ref{vdd}),
formally $O(\Delta^2)$, can be treated in a similar way:\
some can be ignored by assuming the Lagrangian space consistency
relation, some are subdominant in the soft-limit (i.e., they vanish
upon division by $P_v (p)$ and sending ${\bf p} \rightarrow 0$), 
and the dominant
terms are those that involve $\langle v^j_{\bf p} \Delta\rangle$
which give:
\begin{equation}
\label{LHSDelta2result}
\begin{split}
- (2\pi)^3 \delta_D ({\bf p}_1 + {\bf p}_2 + {\bf p})
P_v ({\rm p}, \eta) \,
\int {d^3 {\bf p}_A \over (2\pi)^3} 
\Bigg[ & {\rm p}^j {\rm p}_2^m {D(\eta_1) \over D'(\eta)} {{\bf p}_1
  \cdot {\bf p} \over {\rm p}^2} \langle \tilde\delta_{{\bf p}_1 +
  {\bf p}} (\eta_1) \tilde\delta_{{\bf p}_2 - {\bf p}_A} (\eta_2)
\Delta^m_{{\bf p}_A} (\eta_2) \rangle^{c'} \, + \\
& {\rm p}^j {\rm p}_1^m {D(\eta_1) \over D'(\eta)} {{\bf p}_1 \cdot
  {\bf p} \over {\rm p}^2} \langle \tilde\delta_{{\bf p}_1 + {\bf p} -
  {\bf p}_A} (\eta_1) \tilde\delta_{{\bf p}_2} (\eta_2) \Delta^m_{{\bf
    p}_A} (\eta_1) \rangle^{c'} \, + (1
\leftrightarrow 2)
\Bigg] \, .
\end{split}
\end{equation}
Thus, combining Eqs.\ (\ref{LHSDelta1result}) and
(\ref{LHSDelta2result}), 
$E_L$ of Eq.\ (\ref{Edef}) can be rewritten as:
\begin{equation}
\label{ELexpanded}
\begin{split}
E_L \equiv & \lim_{{\bf p} \rightarrow 0}
{\langle v^j_{{\bf k} = {\bf p}} (\eta) \, {\delta}_{{\bf k}_1 = {\bf
      p}_1} (\eta_1)  \,  {\delta}_{{\bf k}_2 = {\bf p}_2} (\eta_2) \, \rangle^{c'} \over
P_v ({\rm p}, \eta)} \\
= & \, i{\rm p}^j \,
{D(\eta_1) \over D'(\eta)} {{\bf p}_1 \cdot {\bf p} \over {\rm p}^2}
\langle \tilde \delta_{{\bf p}_1} (\eta_1) \tilde
\delta_{{\bf p}_2} (\eta_2)
\rangle^{c'} \, - \int {d^3 {\bf p}_A \over (2\pi)^3} 
\Bigg[ {\rm p}^j {\rm p}_2^m {D(\eta_1) \over D'(\eta)} {{\bf p}_1
  \cdot {\bf p} \over {\rm p}^2} \langle \tilde\delta_{{\bf p}_1} (\eta_1) \tilde\delta_{{\bf p}_2 - {\bf p}_A} (\eta_2)
\Delta^m_{{\bf p}_A} (\eta_2) \rangle^{c'} \\ & +
 {\rm p}^j {\rm p}_1^m {D(\eta_1) \over D'(\eta)} {{\bf p}_1 \cdot
  {\bf p} \over {\rm p}^2} \langle \tilde\delta_{{\bf p}_1 -
  {\bf p}_A} (\eta_1) \tilde\delta_{{\bf p}_2} (\eta_2) \Delta^m_{{\bf
    p}_A} (\eta_1) \rangle^{c'} \, 
\Bigg] + (1 \leftrightarrow 2) + ... \, .
\end{split}
\end{equation}

Next, let us rewrite $E_R$ using the same strategy:
\begin{equation}
\label{ERexpanded}
\begin{split}
E_R \equiv &
- {i {\rm p}^j} \sum_{a=1}^2 {D(\eta_a) \over D'(\eta)} 
{{\bf p} \cdot {{\bf p}_a} \over {\rm p}^2} \langle {\delta}_{{\bf
    k}_1 = {\bf p}_1}  (\eta_1) \, {\delta}_{{\bf k}_2 = {\bf p}_2} (\eta_2) \,
\rangle^{c'} \\
= & - {i {\rm p}^j}  {D(\eta_1) \over D'(\eta)} 
{{\bf p} \cdot {{\bf p}_1} \over {\rm p}^2} \langle \tilde
\delta_{{\bf p}_1}  (\eta_1) \, \tilde \delta_{{\bf p}_2} (\eta_2) \,
\rangle^{c'} \\
& + \int {d^3 {\bf p}_A \over (2\pi)^3} 
\Bigg[ {\rm p}^j {\rm p}_2^m {D(\eta_1) \over D'(\eta)} {{\bf p}_1
  \cdot {\bf p} \over {\rm p}^2} \langle \tilde\delta_{{\bf p}_1} (\eta_1) \tilde\delta_{{\bf p}_2 - {\bf p}_A} (\eta_2)
\Delta^m_{{\bf p}_A} (\eta_2) \rangle^{c'} \\ & +
 {\rm p}^j {\rm p}_1^m {D(\eta_1) \over D'(\eta)} {{\bf p}_1 \cdot
  {\bf p} \over {\rm p}^2} \langle \tilde\delta_{{\bf p}_1 -
  {\bf p}_A} (\eta_1) \tilde\delta_{{\bf p}_2} (\eta_2) \Delta^m_{{\bf
    p}_A} (\eta_1) \rangle^{c'} \, 
\Bigg] + (1 \leftrightarrow 2) + ... \, .
\end{split}
\end{equation}
Thus, we see that $E \equiv E_L + E_R = 0$, 
at least to the two lowest non-trivial orders in
$\Delta$. The cancelation works like this:\ expanding
$e^{i{\bf p}_1 \cdot \pmb{\Delta}}$ and $e^{i{\bf p}_2 \cdot
  \pmb{\Delta}}$ as a formal power series in $\Delta$, a given order
for $E_L$ is canceled by one lower order for $E_R$. 
It can be shown that this pattern continues to arbitrarily high
orders. The proof is given in the Appendix.
This completes our derivation of the Eulerian space consistency
relation, embodied in the statement $E = 0$ (Eq.\ \ref{Edef}), 
from the Lagrangian space consistency relation
(Eq.\ \ref{KRPPLag00repeat}).

\section{Discussion}%
\label{discuss}

We have shown that the consistency relation takes a particularly
simple form in Lagrangian space:\ the squeezed correlation function,
suitably normalized, vanishes (Eqs.\ \ref{KRPPLag}):
\begin{eqnarray}
\label{KRPPdiscuss}
\lim_{{\bf p} \rightarrow 0}
{\langle \pmb{\Delta}_{\bf p} (\eta) \, {\cal O}_{{\bf p}_1} (\eta_1)
  \, ... \, {\cal O}_{{\bf
      p}_N} (\eta_N) \, \rangle^{c'} \over P_\Delta ({\rm p}, \eta)} = 0 \, ,
\end{eqnarray}
where $\Delta$ is the displacement, and ${\cal O}$ can be many
different observables such as mass or galaxy density; the quantities
can be at different times, and ${\bf p}, {\bf p}_1, {\bf p}_2, ...$
label the momenta with ${\bf p}$ being the soft one.\footnote{See also Eq.\ \ref{KRPPLag00} with velocity $\pmb{v}$ as the soft mode.}
The derivation given in \S \ref{derivation} is fully non-perturbative
and is valid even in the presence of multiple-streaming. 
It makes use of a master formula that was derived in an earlier paper
\cite{us}, which relates an (N+1)-point function to
the linear transformation of an N-point function,
for a general nonlinearly-realized symmetry
(Eq.\ \ref{consistencymasterrelation}).
The key realization is that
the nonlinearly-realized symmetry of interest --
the displacement symmetry -- does not
require transforming the Lagrangian coordinate ${\bf q}$
(Eq.\ \ref{symmetryLag}):
\begin{eqnarray}
\label{symmetryLagrepeat}
{\bf q} \rightarrow {\bf q} \quad , \quad
\pmb{\Delta} \rightarrow \pmb{\Delta} + {\bf n}(\eta) \quad , \quad
\Phi \rightarrow \Phi - \left( {\bf n}'' + {a' \over a} {\bf n}'
\right) \cdot {\bf x} \, ,
\end{eqnarray}
where $\pmb{\Delta}$ is the displacement,
${\bf n}$ is some function of time,
$\Phi$ is the gravitational potential, $a$ is the scale factor and
${\bf x}$ is the Eulerian coordinate.\footnote{If there are multiple species present such as
dark matter and galaxies, the same transformation applies to the displacement
of all species. See Eq.\ (\ref{symmetryLag2}).}
Thus, many observables ${\cal O}$ such as the mass density or the galaxy
density,\footnote{See footnote \ref{densitydef}.} when expressed as functions of the Lagrangian
coordinate ${\bf q}$, do not receive linear transformations, and so 
the right hand side of the master formula vanishes.\footnote{Note that even quantities such as $\Delta$ or $\Phi$ have
no linear shift (a shift that is linear in fluctuation variables). More complicated observables could have
a linear shift; see footnote \ref{linearshift}.}
This contrasts with what happens when these observables
are thought of as functions of the Eulerian coordinate ${\bf x}$; they
receive linear transformations because under the same symmetry
\begin{eqnarray}
{\bf x} \rightarrow {\bf x} + {\bf n} (\eta) \, .
\end{eqnarray}

It is worth mentioning that the observables ${\cal O}$ can even be
quantities in redshift space.
For instance, the redshift space mass density $\delta_s ({\bf s})$,
where ${\bf s}$ is the redshift space coordinate, can always be written
as a function of ${\bf q}$, just like what we have done
for the real (Eulerian) space mass density.
This works even in the presence of (real or redshift-space)
multiple-streaming.

It is worth reviewing the assumptions behind the master formula:\
it assumes (1) single-field/clock initial condition -- that which
follows from single-field inflation -- and (2) the adiabatic mode
condition (first emphasized in \cite{Weinberg:2003sw}) -- that the (displacement)
mode generated by the symmetry of interest is the long wavelength
limit of an actual physical mode. In particular, the latter condition
requires that ${\bf n}(\eta)$ have the same time dependence as 
the linear growth factor $D(\eta)$.\footnote{In the presence of multiple species, such as dark matter
and galaxies, the fact that the displacements for all species
have to be transformed by the same ${\bf n}(\eta) \propto D(\eta)$
is a manifestation of the equivalence principle:\ all species
fall at the same rate on large scales.}
This is the reason why 
the growth factor shows up on the right hand side of the
Eulerian space consistency relation (Eq.\ \ref{KRPPEuler}). 
It is interesting that because the Lagrangian space consistency
relation has a vanishing right hand side, the time-dependence
of ${\bf n}(\eta)$ has no direct bearing on the form it takes.

An important point:\
the fact that the consistency relation can be written in such a
simple, even trivial, form in Lagrangian space should not
be taken to imply the lack of physical meaning. Indeed, the consistency
relation can be violated if the initial conditions were not
of the single-field/clock type.
Rather, the simplicity 
suggests that an analytic understanding of nonlinear clustering
is perhaps more promising in Lagrangian space.
This view has a long history, starting from Zeldovich \cite{Zeldovich} (see also \cite{Porto:2013qua, Senatore:2014via}).
What is interesting is that the consistency relation, by virtue of
its being a symmetry statement, is non-perturbative, and thus goes beyond
perturbative treatments such as the Zeldovich approximation.

Ultimately, observations are performed in Eulerian space, not
Lagrangian space. At the nonlinear level, the relation between the
two descriptions is complex. Our derivation of the consistency relation
in Eulerian space from its counterpart in
Lagrangian space is a case in point (\S \ref{nonPTcheck}).
It requires a formal series expansion in the displacement $\Delta$.
In relating the two descriptions, the expansion in
$\Delta$ is done in an uneven manner:\ only phase factors 
such as $e^{i {\bf p}_1 \cdot \pmb{\Delta}}$ are expanded even
though other variables, such as the density $\delta$, also depend on
the displacement. This is not unexpected in relations that are
purported to be non-perturbative -- {\it partial} resummation of
perturbations is often a useful technique.
Can our example point to a useful, new resummation scheme?

A natural question is whether
there are relativistic generalizations
of statements like 
Eq.\ (\ref{KRPPdiscuss}) -- consistency relations
with a vanishing right hand side.
The Lagrangian coordinate (attached to dark matter particles) is essentially the freely-falling
coordinate. Indeed \cite{Pajer} showed that using the freely-falling
coordinate, the dilation consistency relation \cite{Maldacena:2002vr} can
be rewritten in a similarly simple form (see also \cite{Tanaka}).
Their derivation is perturbative. It should be possible
to extend their proof using the non-perturbative arguments
presented here. More generally, it would be interesting to see if
further general relativistic consistency relations, such as those
found by \cite{Hinterbichler:2013dpa},  can also be recast in this fashion.

\section*{Acknowledgements}
We thank Paolo Creminelli and Donghui Jeong for useful discussions.
This work is supported in part by the United States Department of
Energy under DOE grant DE-FG02-92-ER40699 and DOE grant DE-SC0011941,
and by NASA under NASA ATP grant NNX10AN14G.

\appendix

\section{Recovering the Eulerian space consistency relation from
  Lagrangian space -- to arbitrary orders in displacement}

In \S \ref{nonPTcheck}, we argue that the Eulerian space consistency
relation follows from its Lagrangian space counter-part, 
at least to the two lowest non-trivial orders in a formal expansion in 
displacement. In this Appendix, we show that this works to arbitrary
orders.

We begin by expanding Eq.\ (\ref{deltakrewrite}) to all orders in $\Delta$:
\begin{equation}
\begin{split}
\delta_{{\bf k}_1 = {\bf p}_1} &= \int d^3 {\bf q} \, J({\bf q})\, \delta({\bf x}({\bf q}))\,e^{i {\bf p}_1 \cdot ({\bf q}+\pmb{\Delta}({\bf q}))}\\
&= \sum_{n=0}^{\infty}\frac{i^n}{n!}{\rm p}_1^{i_1}\cdots {\rm p}_1^{i_n}\int_{{\bf p}_{j_1}, \cdots ,{\bf p}_{j_n}} \tilde{\delta}_{{\bf p}_1 - {\bf p}_{j_1} - \cdots - {\bf p}_{j_n}}\Delta^{i_1}_{{\bf p}_{j_1}} \cdots \Delta^{i_n}_{{\bf p}_{j_n}}
\end{split}
\end{equation}
where as before, $\tilde{\delta}({\bf q}) = J({\bf q})\delta({\bf x}(\bf q))$.  Collecting together the terms of order $\Delta^n$ in the three-point correlator $\langle v^j_{\bf p}(\eta) \delta_{{\bf p}_1}(\eta_1) \delta_{{\bf p}_2}(\eta_2)\rangle$ in $E_L$, we have
\begin{equation}
\begin{split}
\sum_{m=0}^{n}\frac{i^m}{m!}\frac{i^{n-m}}{(n-m)!}{\rm p}_1^{i_1}\cdots {\rm p}_1^{i_m}{\rm p}_2^{i_{m+1}}\cdots {\rm p}_2^{i_n} \int_{{\bf p}_{j_1}, \cdots ,{\bf p}_{j_n}}\langle & v^j_{\bf p}(\eta)\tilde{\delta}_{{\bf p}_1 - {\bf p}_{j_1} - \cdots - {\bf p}_{j_m}}(\eta_1)\Delta^{i_1}_{{\bf p}_{j_1}}(\eta_1) \cdots \Delta^{i_m}_{{\bf p}_{j_m}}(\eta_1)\\
&  \cdot \tilde{\delta}_{{\bf p}_2 - {\bf p}_{j_{m+1}} - \cdots - {\bf p}_{j_n}}(\eta_2)\Delta^{i_{m+1}}_{{\bf p}_{j_{m+1}}}(\eta_2) \cdots \Delta^{i_n}_{{\bf p}_{j_n}}(\eta_2)\rangle
\end{split}
\end{equation} 
where the correlator in the integral is the full correlator containing both connected and disconnected pieces, but where no proper subset of the original momenta ${\bf p}, {\bf p}_1, {\bf p}_2$ sums to zero, since it is the connected correlator that appears in $E_L$.  The correlator can be split into a sum over products of connected blocks.  Anticipating division by $P_v({\bf p})$, we see that the Lagrangian space consistency relation implies that all contributions where the soft velocity $v^j({\bf p})$ is part of a connected correlator with two or more other fields will vanish.  The remaining terms contain a factor of either
\begin{equation}
\langle v^j({\bf p}) \tilde{\delta}_{{\bf p}_1 - \cdots }\rangle \, , \, \langle v^j({\bf p}) \tilde{\delta}_{{\bf p}_2 - \cdots }\rangle \, , \, \langle v^j({\bf p}) \Delta_{{\bf p}_j}(\eta_1)\rangle \, , \, \langle v^j({\bf p}) \Delta_{{\bf p}_j}(\eta_2)\rangle \,.
\end{equation}
The first two types of terms will be suppressed by an additional power of the soft momentum, and are subdominant in the squeezed limit.  The final set of terms are
\begin{equation}
\begin{split}
  \sum_{m=0}^{n}\frac{i^m}{m!}\frac{i^{n-m}}{(n-m)!}{\rm p}_1^{i_1}&\cdots {\rm p}_1^{i_m}{\rm p}_2^{i_{m+1}}\cdots {\rm p}_2^{i_n}  \int_{{\bf p}_{j_1}, \cdots ,{\bf p}_{j_{n}}} \Bigg[ m \langle v^j_{\bf p}(\eta)\Delta^{i_1}_{{\bf p}_{j_1}}(\eta_1)\rangle \langle\tilde{\delta}_{{\bf p}_1 - {\bf p}_{j_1} - \cdots - {\bf p}_{j_m}}(\eta_1)\\
&\cdot \Delta^{i_1}_{{\bf p}_{j_1}}(\eta_1) \cdots \Delta^{i_m}_{{\bf p}_{j_m}}(\eta_1)\tilde{\delta}_{{\bf p}_2 - {\bf p}_{j_{m+1}} - \cdots - {\bf p}_{j_n}}(\eta_2)\Delta^{i_{m+1}}_{{\bf p}_{j_{m+1}}}(\eta_2) \cdots \Delta^{i_n}_{{\bf p}_{j_n}}(\eta_2)\rangle \Bigg]\\
& + (1 \leftrightarrow 2)
\end{split}
\end{equation}
and they give (relabeling $m$ as $m + 1$)
\begin{equation}
\begin{split}
  \Bigg[ i {\rm p}^j \frac{{\bf p}\cdot {\bf p}_1}{{\bf p}^2}\frac{D(\eta_1)}{D'(\eta)}\sum_{m=0}^{n-1}&\frac{i^{n-1}}{m!(n-1-m)!}{\rm p}_1^{i_1}\cdots {\rm p}_1^{i_m}{\rm p}_2^{i_{m+1}}\cdots {\rm p}_2^{i_n} \int_{{\bf p}_{j_1}, \cdots ,{\bf p}_{j_{n}}} \langle \tilde{\delta}_{{\bf p}_1 + {\bf p}- {\bf p}_{j_1} - \cdots - {\bf p}_{j_m}}(\eta_1)\\
& \cdot \Delta^{i_1}_{{\bf p}_{j_1}}(\eta_1) \cdots \Delta^{i_m}_{{\bf p}_{j_m}}(\eta_1)\tilde{\delta}_{{\bf p}_2 - {\bf p}_{j_{m+1}} - \cdots - {\bf p}_{j_n}}(\eta_2)\Delta^{i_{m+1}}_{{\bf p}_{j_{m+1}}}(\eta_2) \cdots \Delta^{i_n}_{{\bf p}_{j_n}}(\eta_2)\rangle\Bigg]\\
& + (1 \leftrightarrow 2)
\end{split}
\end{equation}
Comparing this to the order $\Delta^{n-1}$ terms in the expansion of $E_R$,
\begin{equation}
\begin{split}
  \Bigg[-i {\rm p}^j \frac{{\bf p}\cdot {\bf p}_1}{{\bf p}^2}\frac{D(\eta_1)}{D'(\eta)}\sum_{m=0}^{n-1}&\frac{i^{n-1}}{m!(n-1-m)!}{\rm p}_1^{i_1}\cdots {\rm p}_1^{i_m}{\rm p}_2^{i_{m+1}}\cdots {\rm p}_2^{i_n} \int_{{\bf p}_{j_1}, \cdots ,{\bf p}_{j_{n}}} \langle \tilde{\delta}_{{\bf p}_1 - {\bf p}_{j_1} - \cdots - {\bf p}_{j_m}}(\eta_1)\\
& \cdot \Delta^{i_1}_{{\bf p}_{j_1}}(\eta_1) \cdots \Delta^{i_m}_{{\bf p}_{j_m}}(\eta_1)\tilde{\delta}_{{\bf p}_2 - {\bf p}_{j_{m+1}} - \cdots - {\bf p}_{j_n}}(\eta_2)\Delta^{i_{m+1}}_{{\bf p}_{j_{m+1}}}(\eta_2) \cdots \Delta^{i_n}_{{\bf p}_{j_n}}(\eta_2)\rangle\Bigg]\\
& + (1 \leftrightarrow 2)
\end{split}
\end{equation}
the terms cancel in the ${\bf p} \to 0$ limit to give $E_L + E_R = 0$ order by order in the formal expansion in $\Delta$, Q.E.D.



\end{document}